\RequirePackage{lineno}
\documentclass[aps,prl,twocolumn,superscriptaddress,showpacs,preprintnumbers,amsmath,amssymb, nobibnotes]{revtex4-1}

\usepackage{graphicx} 
\usepackage{dcolumn}  
\graphicspath{{ps}}
\usepackage{enumitem}  
\usepackage{feynmf}    
\newcommand{\overbar}[1]{\mkern 1.5mu\overline{\mkern-1.5mu#1\mkern-0.1mu}\mkern 0.1mu}

\newcommand{\bks}{$B^{0} \to K^{0}_{S}K^{0}_{S}K^{0}_{S}$}
\newcommand{\mksks}{$M_{K^{0}_{S}K^{0}_{S}}$}
\newcommand{\cps}{$\mathcal{S}$}
\newcommand{\cpa}{$\mathcal{A}$}
\newcommand{\onn}{$\mathcal{O}_{\rm NN}$}
\newcommand{\tonn}{$\mathcal{O}'_{\rm NN}$}

\begin{document}

\preprint{\vbox{ \hbox{   }
						\hbox{Belle preprint {\it 2020-17}}
						\hbox{KEK preprint {\it 2020-34}}
}}

\title{ \quad\\[1.0cm] Measurement of time-dependent $C\!P$ violation parameters in \bks ~decays at Belle}

\noaffiliation
\affiliation{University of the Basque Country UPV/EHU, 48080 Bilbao}
\affiliation{Beihang University, Beijing 100191}
\affiliation{University of Bonn, 53115 Bonn}
\affiliation{Brookhaven National Laboratory, Upton, New York 11973}
\affiliation{Budker Institute of Nuclear Physics SB RAS, Novosibirsk 630090}
\affiliation{Faculty of Mathematics and Physics, Charles University, 121 16 Prague}
\affiliation{Chonnam National University, Gwangju 61186}
\affiliation{University of Cincinnati, Cincinnati, Ohio 45221}
\affiliation{Deutsches Elektronen--Synchrotron, 22607 Hamburg}
\affiliation{Department of Physics, Fu Jen Catholic University, Taipei 24205}
\affiliation{Key Laboratory of Nuclear Physics and Ion-beam Application (MOE) and Institute of Modern Physics, Fudan University, Shanghai 200443}
\affiliation{Justus-Liebig-Universit\"at Gie\ss{}en, 35392 Gie\ss{}en}
\affiliation{II. Physikalisches Institut, Georg-August-Universit\"at G\"ottingen, 37073 G\"ottingen}
\affiliation{SOKENDAI (The Graduate University for Advanced Studies), Hayama 240-0193}
\affiliation{Gyeongsang National University, Jinju 52828}
\affiliation{Department of Physics and Institute of Natural Sciences, Hanyang University, Seoul 04763}
\affiliation{University of Hawaii, Honolulu, Hawaii 96822}
\affiliation{High Energy Accelerator Research Organization (KEK), Tsukuba 305-0801}
\affiliation{J-PARC Branch, KEK Theory Center, High Energy Accelerator Research Organization (KEK), Tsukuba 305-0801}
\affiliation{Higher School of Economics (HSE), Moscow 101000}
\affiliation{Forschungszentrum J\"{u}lich, 52425 J\"{u}lich}
\affiliation{IKERBASQUE, Basque Foundation for Science, 48013 Bilbao}
\affiliation{Indian Institute of Science Education and Research Mohali, SAS Nagar, 140306}
\affiliation{Indian Institute of Technology Bhubaneswar, Satya Nagar 751007}
\affiliation{Indian Institute of Technology Hyderabad, Telangana 502285}
\affiliation{Indian Institute of Technology Madras, Chennai 600036}
\affiliation{Indiana University, Bloomington, Indiana 47408}
\affiliation{Institute of High Energy Physics, Chinese Academy of Sciences, Beijing 100049}
\affiliation{Institute of High Energy Physics, Vienna 1050}
\affiliation{Institute for High Energy Physics, Protvino 142281}
\affiliation{INFN - Sezione di Napoli, 80126 Napoli}
\affiliation{INFN - Sezione di Torino, 10125 Torino}
\affiliation{Advanced Science Research Center, Japan Atomic Energy Agency, Naka 319-1195}
\affiliation{J. Stefan Institute, 1000 Ljubljana}
\affiliation{Institut f\"ur Experimentelle Teilchenphysik, Karlsruher Institut f\"ur Technologie, 76131 Karlsruhe}
\affiliation{Kavli Institute for the Physics and Mathematics of the Universe (WPI), University of Tokyo, Kashiwa 277-8583}
\affiliation{Kennesaw State University, Kennesaw, Georgia 30144}
\affiliation{Department of Physics, Faculty of Science, King Abdulaziz University, Jeddah 21589}
\affiliation{Kitasato University, Sagamihara 252-0373}
\affiliation{Korea Institute of Science and Technology Information, Daejeon 34141}
\affiliation{Korea University, Seoul 02841}
\affiliation{Kyungpook National University, Daegu 41566}
\affiliation{LAL, Univ. Paris-Sud, CNRS/IN2P3, Universit\'{e} Paris-Saclay, Orsay 91898}
\affiliation{P.N. Lebedev Physical Institute of the Russian Academy of Sciences, Moscow 119991}
\affiliation{Faculty of Mathematics and Physics, University of Ljubljana, 1000 Ljubljana}
\affiliation{Ludwig Maximilians University, 80539 Munich}
\affiliation{Luther College, Decorah, Iowa 52101}
\affiliation{University of Maribor, 2000 Maribor}
\affiliation{Max-Planck-Institut f\"ur Physik, 80805 M\"unchen}
\affiliation{School of Physics, University of Melbourne, Victoria 3010}
\affiliation{University of Mississippi, University, Mississippi 38677}
\affiliation{University of Miyazaki, Miyazaki 889-2192}
\affiliation{Moscow Physical Engineering Institute, Moscow 115409}
\affiliation{Graduate School of Science, Nagoya University, Nagoya 464-8602}
\affiliation{Universit\`{a} di Napoli Federico II, 80055 Napoli}
\affiliation{Nara Women's University, Nara 630-8506}
\affiliation{National Central University, Chung-li 32054}
\affiliation{National United University, Miao Li 36003}
\affiliation{Department of Physics, National Taiwan University, Taipei 10617}
\affiliation{H. Niewodniczanski Institute of Nuclear Physics, Krakow 31-342}
\affiliation{Nippon Dental University, Niigata 951-8580}
\affiliation{Niigata University, Niigata 950-2181}
\affiliation{University of Nova Gorica, 5000 Nova Gorica}
\affiliation{Novosibirsk State University, Novosibirsk 630090}
\affiliation{Okinawa Institute of Science and Technology, Okinawa 904-0495}
\affiliation{Osaka City University, Osaka 558-8585}
\affiliation{Pacific Northwest National Laboratory, Richland, Washington 99352}
\affiliation{Panjab University, Chandigarh 160014}
\affiliation{Peking University, Beijing 100871}
\affiliation{University of Pittsburgh, Pittsburgh, Pennsylvania 15260}
\affiliation{Punjab Agricultural University, Ludhiana 141004}
\affiliation{Theoretical Research Division, Nishina Center, RIKEN, Saitama 351-0198}
\affiliation{University of Science and Technology of China, Hefei 230026}
\affiliation{Seoul National University, Seoul 08826}
\affiliation{Showa Pharmaceutical University, Tokyo 194-8543}
\affiliation{Soochow University, Suzhou 215006}
\affiliation{Soongsil University, Seoul 06978}
\affiliation{Sungkyunkwan University, Suwon 16419}
\affiliation{School of Physics, University of Sydney, New South Wales 2006}
\affiliation{Department of Physics, Faculty of Science, University of Tabuk, Tabuk 71451}
\affiliation{Tata Institute of Fundamental Research, Mumbai 400005}
\affiliation{Department of Physics, Technische Universit\"at M\"unchen, 85748 Garching}
\affiliation{School of Physics and Astronomy, Tel Aviv University, Tel Aviv 69978}
\affiliation{Toho University, Funabashi 274-8510}
\affiliation{Department of Physics, Tohoku University, Sendai 980-8578}
\affiliation{Earthquake Research Institute, University of Tokyo, Tokyo 113-0032}
\affiliation{Department of Physics, University of Tokyo, Tokyo 113-0033}
\affiliation{Tokyo Institute of Technology, Tokyo 152-8550}
\affiliation{Tokyo Metropolitan University, Tokyo 192-0397}
\affiliation{Virginia Polytechnic Institute and State University, Blacksburg, Virginia 24061}
\affiliation{Wayne State University, Detroit, Michigan 48202}
\affiliation{Yamagata University, Yamagata 990-8560}
\affiliation{Yonsei University, Seoul 03722}
  \author{K.~H.~Kang}\affiliation{Kyungpook National University, Daegu 41566} 
  \author{H.~Park}\affiliation{Kyungpook National University, Daegu 41566} 
  \author{T.~Higuchi}\affiliation{Kavli Institute for the Physics and Mathematics of the Universe (WPI), University of Tokyo, Kashiwa 277-8583} 
  \author{K.~Miyabayashi}\affiliation{Nara Women's University, Nara 630-8506} 
  \author{K.~Sumisawa}\affiliation{High Energy Accelerator Research Organization (KEK), Tsukuba 305-0801}\affiliation{SOKENDAI (The Graduate University for Advanced Studies), Hayama 240-0193} 
  \author{I.~Adachi}\affiliation{High Energy Accelerator Research Organization (KEK), Tsukuba 305-0801}\affiliation{SOKENDAI (The Graduate University for Advanced Studies), Hayama 240-0193} 
  \author{J.~K.~Ahn}\affiliation{Korea University, Seoul 02841} 
  \author{H.~Aihara}\affiliation{Department of Physics, University of Tokyo, Tokyo 113-0033} 
  \author{S.~Al~Said}\affiliation{Department of Physics, Faculty of Science, University of Tabuk, Tabuk 71451}\affiliation{Department of Physics, Faculty of Science, King Abdulaziz University, Jeddah 21589} 
  \author{D.~M.~Asner}\affiliation{Brookhaven National Laboratory, Upton, New York 11973} 
% ...
\collaboration{The Belle Collaboration}


\begin{abstract}
We measure the time-dependent $C\!P$ violation parameters in \bks ~decays using $772\times 10^6 ~ B\overbar{B}$ pairs collected at the $\Upsilon (4S)$ resonance with the Belle detector at the KEKB asymmetric-energy $e^+ e^-$ collider.
The obtained mixing-induced and direct $C\!P$ asymmetries are 
 $-0.71 \pm 0.23~{\rm(stat)} \pm 0.05~{\rm(syst)}$ 
 and $0.12 \pm 0.16~{\rm(stat)} \pm 0.05~{\rm(syst)}$, respectively.
These values are consistent with the Standard Model predictions.
The significance of $C\!P$ violation differs from zero by 2.5 standard deviations.

\end{abstract}

\pacs{13.20.He, 14.40.Nd}

\maketitle
\begin{figure}[htb]
\includegraphics[width=0.44\textwidth]{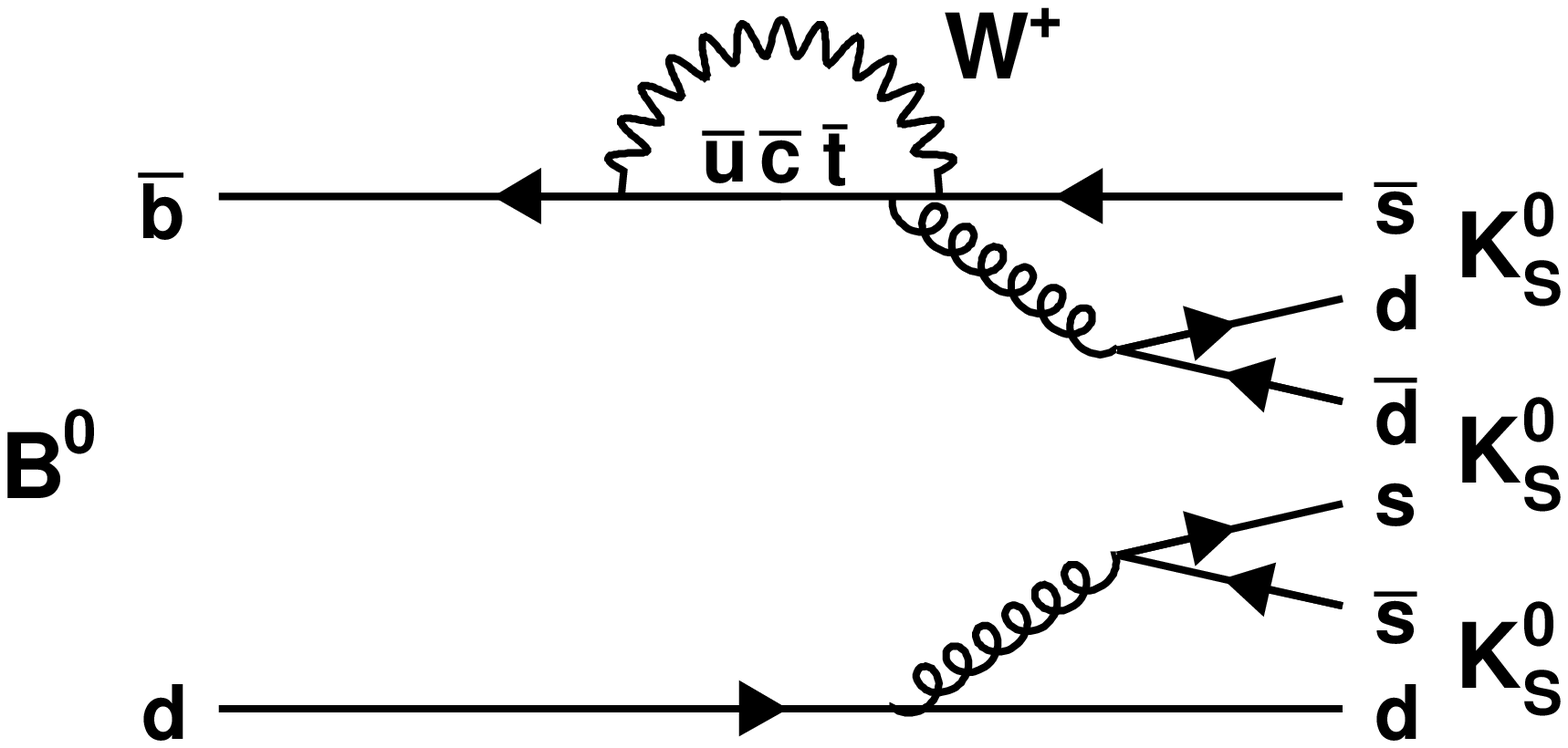}
\caption{
Penguin amplitude for the \bks ~decays.
}
\label{fig_feynman}
\end{figure}

\tighten
{\renewcommand{\thefootnote}{\fnsymbol{footnote}}}
\setcounter{footnote}{0}
In the Standard Model (SM), $C\!P$ violation in the quark sector is described by an irreducible phase in the Kobayashi-Maskawa (KM) mechanism~\cite{KM}.
The charmless three-body decay \bks ~is mediated by the $b \to sq\overbar{q}$ quark transition, which is prohibited in the lowest-order SM interaction.
Instead, this $C\!P$-even decay occurs via a ``penguin'' amplitude, as shown in Fig.~\ref{fig_feynman}.
Deviations from the SM expectations for $C\!P$-violating parameters provide sensitivity to new physics~\cite{np1}.

Time-dependent $C\!P$ violation can be caused by interference between the decay and mixing amplitudes.
When one of the neutral $B$ mesons produced from the $\Upsilon (4S)$ decays
 into a $C\!P$ eigenstate, $f_{C\!P}$, at time $t_{C\!P}$,
 and the other into a flavor-distinguishable final state,
 $f_{\rm tag}$, at time $t_{\rm tag}$, the time-dependent decay rate is given by~\cite{Sanda}

\begin{equation} \label{eqn_dt1}
\begin{split}
\mathcal{P}(\Delta t) & =   \frac{e^{-|\Delta t|/\tau_{B^0}}}{4\tau_{B^0}}\times \\ 
 &  (1 + q[ \mathcal{S}\sin(\Delta m^{}_d\Delta t) 
+ \mathcal{A}\cos(\Delta m^{}_d\Delta t)  ] ),
\end{split}
\end{equation}

 where $\Delta t \equiv t_{C\!P} - t_{\rm tag}$,
 measured in the center-of-mass (CM) frame,
 and the $C\!P$-violating parameters \cps ~and \cpa ~are related to mixing-induced and direct $C\!P$ violation, respectively.
Here the flavor $q$ is $+1$ ($-1$) for $\overbar{B}^0$ ($B^0$),
 $\tau_{B^0}$ is the $B^0$ lifetime,
 and $\Delta m_d$ is the mass difference between the two mass eigenstates of the $B^0$-$\overbar{B}^0$ system.
The SM predicts that $\mathcal{S} = -\sin 2 \phi_1$ and $\mathcal{A} = 0$ in \bks, where $\phi_1 \equiv \arg [-V_{cd}V_{cb}^*/V_{td}V_{tb}^*]$ \cite{phi1notation}.

Previous measurements of \cps ~at Belle and BaBar have yielded values of $-0.30\pm 0.32~{\rm (stat)}\pm 0.08~{\rm (syst)}$ using $535 \times 10^6$ $B\overbar{B}$ pairs,
 and $-0.94^{+0.24}_{-0.21}~{\rm (stat)}\pm 0.06~{\rm (syst)}$ using $468 \times 10^6$ $B\overbar{B}$ pairs, respectively~\cite{b03ksbelle, b03ksbabar}.
To search for physics beyond the SM containing a new $C\!P$-violating phase,
 we measure \cps~and \cpa~in \bks ~decays with the final Belle data set
 of $772 \times 10^6$ $B\overbar{B}$ pairs.

The Belle detector is a large-solid-angle magnetic spectrometer
 that consists of a silicon vertex detector (SVD),
 a 50-layer central drift chamber (CDC),
 an array of aerogel threshold Cherenkov counters (ACC),
 a barrel-like arrangement of time-of-flight scintillation counters (TOF),
 and an electromagnetic calorimeter comprised of CsI(Tl) crystals (ECL).
 These detector components are located inside a superconducting solenoid coil
 that provides a 1.5~T magnetic field.
An iron flux-return located outside the magnetic coil is instrumented 
 to detect $K_L^0$ mesons and identify muons (KLM).
The detector is described in detail elsewhere~\cite{Belle}.
Two inner detector configurations were used.
A 2.0 cm radius beampipe with a double-wall beryllium structure
 and a three-layer SVD were used for the first sample of $152 \times 10^6 ~ B\overbar{B}$ pairs,
 while a 1.5 cm radius beampipe, a four-layer SVD
 and a small-inner-cell CDC were used to record the remaining $620 \times 10^6 ~ B\overbar{B}$ pairs~\cite{svd2}.
The latter data sample has been reprocessed with improved software, which incorporates an improved vertex reconstruction~\cite{pbf, ccbar}. 


The $\Upsilon (4S)$ is produced at the KEKB asymmetric-energy $e^+ e^-$ collider~\cite{KEKB} with
 a Lorentz boost (${\beta \gamma}$) of 0.425; it subsequently decays to $B$ and $\overbar{B}$ mesons,
 which are nearly at rest in the CM frame.
The Lorentz boost introduces a sufficient distance between the $B$ and $\overbar{B}$ decay vertices to be measurable 
 nearly along with the $z$ axis, which is antiparallel to the $e^+$ beam direction.
The distance is related to $\Delta t \approx (z_{C\!P} - z_{\rm tag}) / c \beta \gamma$, where $z_{C\!P}$ and $z_{\rm tag}$ are the coordinates of the decay positions of $f_{C\!P}$ and $f_{\rm tag}$, respectively.
To avoid the large backgrounds accompanying $\gamma$ and $\pi^0$ detection,
 we reconstruct the $K_S^0$ only through its decay to two charged pions.
The event selection and measurement of $C\!P$ violation parameters are optimized using Monte Carlo (MC) events.
The MC events are generated by E{\scriptsize VT}G{\scriptsize EN}~\cite{Evtgen}, and the detector response is modeled using G{\scriptsize EANT}3~\cite{GEANT3}.
We simulate the $B$-meson decay to three $K_S^0$ as uniformly distributed in the available phase space.


The $K_S^0$ is selected from charged pion pairs using a neural network (NN) ~\cite{nisks,NB} with 13 inputs:
 the $K_S^0$ momentum in the lab frame ($>$ 0.06 GeV/$c$);
 the distance between the pion tracks in the $z$-direction ($<$ 20 cm); 
 the flight length in the $x$-$y$ plane; 
 the angle between the $K_S^0$ momentum and the vertex displacement vector;
 the angle between the $K_S^0$ momentum and the pion momentum; 
 for each daughter pion, 
 the distance of closest approach to the interaction point (IP); 
 the existence of the SVD hits; and 
 the number of axial- and stereo-wire hits in the CDC.
The mass ranges allowed are 0.474 GeV/$c^{2}$ $< M(\pi^+ \pi^-) <$ 0.522 GeV/$c^{2}$ when only one pion hits the SVD,
 and 0.478 GeV/$c^{2}$ $< M(\pi^+ \pi^-) <$ 0.517 GeV/$c^{2}$ otherwise. 

To identify the signal $B$-decay, we use the energy difference $\Delta E \equiv E_{\rm beam} - E_{B}$
 and the beam-energy-constrained mass ${M_{\rm bc} \equiv \sqrt{(E_{\rm beam})^2 - |\vec{p}_B|^{2} c^{2}}/c^2}$,
 where $E_{\rm beam}$ is the beam energy, and 
$E_{B}$ and $\vec{p}_{B}$ are the energy and momentum, respectively, of the $B^0$ candidate.
All quantities are evaluated in the CM frame.
The $B^0$ candidates are required to lie in the region of $M_{\rm bc} > 5.2$ GeV$/c^2$ and $|\Delta E| < 0.2$ GeV.

We find that seven percent of the events have more than one $B^0$ candidate.
When there are multiple $B^0$ candidates in an event, 
 we choose the one with the smallest $\chi^2$ as defined by 
$\chi^2 = \sum_{i=1}^{3}  [(M_i(\pi^{+}\pi^{-}) - m_{K^{0}_{S}}) / \sigma_{i}]^2$, 
 where $M_i(\pi^+\pi^-)$ and $\sigma_i$ are the invariant mass and mass resolution for the $i$-th $K^0_S$, respectively, and $m_{K^0_S}$ is the nominal $K^0_S$ mass~\cite{PDG}.


The dominant source of background is continuum $e^+ e^- \to q\overbar{q} ~(q= u, d, s, c)$ events.
To suppress this background, we use another NN with the following inputs:
 the cosine of the polar angle of the $B^0$-candidate flight direction in the CM frame ($\cos \theta_B$);
 the cosine of the angle between the thrust axis of the $B^0$ candidate and that of the rest of the event ($\cos \theta_T$); and
a likelihood ratio obtained from modified Fox-Wolfram moments~\cite{KSFW}.
The NN outputs (\onn) range between $+1$ and $-1$, where \onn~ close to $+1$ ($-1$) indicates a signal-like (background-like) event.
The \onn~criterion is obtained by maximizing a figure-of-merit (FOM = $N_{\rm sig}/\sqrt{N_{\rm sig}+N_{q\overbar{q}}}$),
 where $N_{\rm sig}$ and $N_{q\overbar{q}}$ are the number of signal and continuum MC events.
The FOM is maximal at the value of \onn~= 0.08.
The signal region is defined as $M_{\rm bc} > 5.27$ GeV$/c^2$ and $|\Delta E| < 0.1$ GeV.
The \onn~ requirement retains 81\% of the signal and reduces continuum background by a factor of 10 in the signal region.

The \onn~is transformed by using the formula
\begin{equation}\label{eqn_dt2}
{\mathcal{O}'_{\rm NN}} = \log( \frac{{\mathcal{O}_{\rm NN} ~ - ~ \mathcal{O}_{\rm NN,min}} } {{\mathcal{O}_{\rm NN,max} ~ - ~ \mathcal{O}_{\rm NN}}}).
\end{equation}
The value of $\mathcal{O}_{\rm NN,min}$ is selected to be 0.08, thus maximizing the FOM. 
The value of $\mathcal{O}_{\rm NN,max}$ is set to 0.99, the highest observed value of \onn.




Among the \bks ~decays, there are quasi-two-body intermediate states, 
 where both $b \to c$ and $b \to s$ transitions contribute.
The former contaminates the measurement of $C\!P$ violation in $b \to s$ transitions.
Among possible $b \to c$ transition-induced $B$ decays,
 we expect significant contributions solely from $B^0 \to \chi_{c0}(\chi_{c0} \to K^{0}_{S}K^{0}_{S})K^{0}_{S}$ decays; this contribution is estimated to be $16.3 \pm 3.1$ events.
We use the invariant mass, \mksks, to veto $B^0 \to \chi_{c0} K^0_S$ decays:
 we reject the $B^0$ candidate if any $K^{0}_{S}$ pair among its decay products has
 an invariant mass within $\pm 2 \sigma$ of the nominal mass,
 where $\sigma$ is the reconstructed $\chi_{c0}$ mass resolution from simulation.
The veto range is 3.388 GeV/$c^2$ $<$ \mksks ~$<$ 3.444 GeV/$c^2$,
 and this removes 83\% of the $B^0 \to \chi_{c0} K^0_S$ decays.


To identify the $B$ meson flavor, a flavor tagging algorithm \cite{TaggingNIM} is used that
 utilizes inclusive properties of particles not associated with the signal decay. 
This algorithm returns the value of $q$ (defined earlier) and a tagging quality variable $r$.
The latter varies from $r=0$ for no tagging information to $r=1$ for unambiguous flavor assignment.
The probability density function (PDF) for signal events modifies Eq.~(\ref{eqn_dt1}) by taking the wrong-tag fraction, $w$, and
 its difference between $B^0$ and $\overbar{B}^0$, $\Delta w$, into account:

\begin{equation}\label{eqn_dt3}
\begin{split}
\mathcal{P_{\rm sig}}(\Delta t,q) & =  \frac{e^{-|\Delta t|/\tau_{B^0}}}{4\tau_{B^0}}  
 \left(\mathstrut^{\mathstrut}_{\mathstrut} 1 + q\Delta w+ (1-2w)q  \right. \\
   &  \left. \times[ \mathcal{S}\sin(\Delta m^{}_d\Delta t) + \mathcal{A}\cos(\Delta m^{}_d\Delta t)  ] \mathstrut^{\mathstrut}_{\mathstrut}\right).
\end{split}
\end{equation}

The events are categorized into seven $r$ bins.
For each of these bins, $w$ and $\Delta w$ are determined by high statistics flavor-specific $B$ meson decays~\cite{ccbar}. 


The parameter $\Delta t$ in Eq.~(\ref{eqn_dt3}) is determined through 
 vertex reconstruction for the signal $B^0$ ($B_{C\!P}$)
 and the accompanying $B^0$ ($B_{\rm tag}$).
For reconstruction of the $B_{C\!P}$ vertex,
we use those $K_{S}^{0}$ trajectories in which both daughter pions have SVD hits in the $z$ direction.
$B_{C\!P}$ can have up to three $K^0_S$ that satisfy this requirement.
According to the signal MC, 14\% of events do not have any $K^0_S$ producing sufficient SVD hits.
The IP information is incorporated as a virtual straight track along the $z$ axis,
 called the ``IP tube'', to provide a constraint for kinematical fits to reconstruct the $B$ decay vertex.
The $B_{C\!P}$ vertex is obtained from the available $K^0_S$ trajectories and the IP tube.
The IP tube is also used to reconstruct the $B_{\rm tag}$ vertex using the charged tracks 
 not assigned to $B_{C\!P}$, as described in more detail in Ref.~\cite{vertexres}.
Because of this treatment, we can reconstruct the $B_{C\!P}$ ($B_{\rm tag}$) vertex with only one $K^0_S$ trajectory (charged track).

Events with poorly reconstructed vertices are rejected by requiring: 

\begin{enumerate}[label=(\arabic*)]
\item $|\Delta t| < 70$ ps; 
\item a vertex quality for each of $B_{C\!P}$ and $B_{\rm tag}$ of less than 50, when the vertex is constrained by multiple tracks; and
\item uncertainties on the $z$ position of the vertices for both $B_{C\!P}$ and $B_{\rm tag}$ of less than 0.2 mm
 when the vertex is constrained by multiple tracks, and less than 0.5 mm 
 when the $B_{C\!P}$ ($B_{\rm tag}$) vertex is constrained by a single $K^0_S$ trajectory (single track).
\end{enumerate}
The vertex quality is $\chi^2$ per degree of freedom,
 where $\chi^2$ is obtained from the vertex fit without the IP tube constraint.
After the poorly reconstructed events are discarded,
 the remaining data events amount to 73\% of the total number of events in the signal region.

\begin{figure}[htb]
\includegraphics[width=0.44\textwidth]{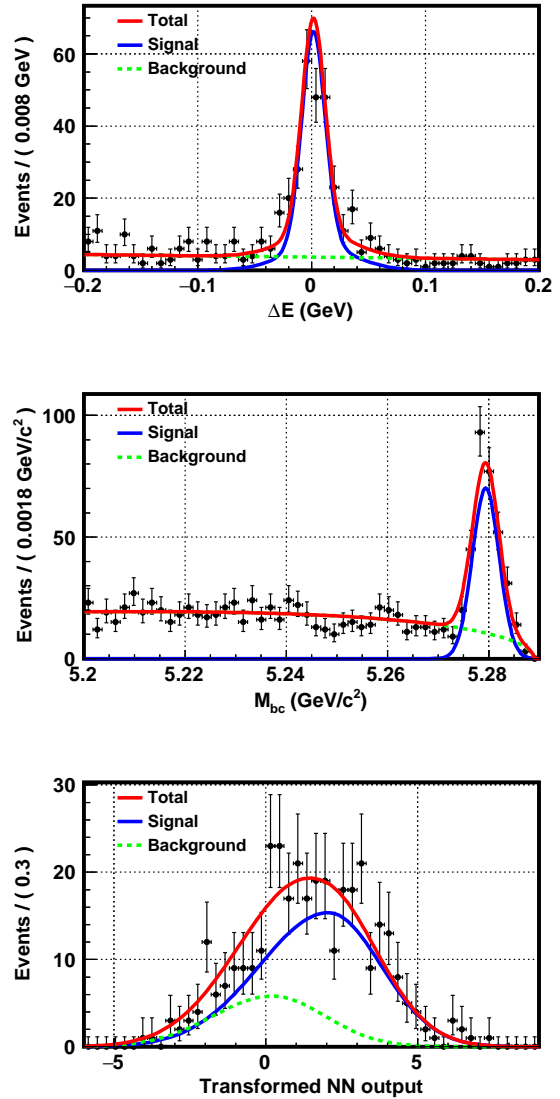}
\caption{
Result of signal-extraction fit to $\Delta E$ (top) in the redefined signal region of $M_{\rm bc}$ and \tonn,
 $M_{\rm bc}$ (middle) in the redefined signal region of $\Delta E$ and \tonn,
 and the \tonn~(bottom) in the redefined signal region of $\Delta E$ and $M_{\rm bc}$ distributions.
The red, blue, and green dashed lines represent the total, signal, and background PDFs.
The points with error bars represent data.
}
\label{fig_sigext}
\end{figure}

We determine the signal yield by performing an extended unbinned maximum-likelihood fit.
The fit is done in the $\Delta E$-$M_{\rm bc}$-\tonn~three-dimensional space,
 with each PDF expressed as the product of the one-dimensional PDFs.
The signal PDFs are modeled as a double Gaussian, a Gaussian, and an asymmetric Gaussian, respectively, 
 and the background PDFs are modeled as a first-order polynomial, an ARGUS function~\cite{ARGUS}, and an asymmetric Gaussian, respectively.
The parameters of the signal PDF are fixed according to fits to the signal MC;
 the parameters of the background PDF are left free in the fit.
The fit result is shown in Fig.~\ref{fig_sigext}.
In the signal region, the signal yield is 258 $\pm$ 17 events and the purity is 74\%,
 where $-4.72 <$ \tonn~ $<7.24$ is further required for the signal region.

\begin{figure}[htb]
\includegraphics[width=0.44\textwidth]{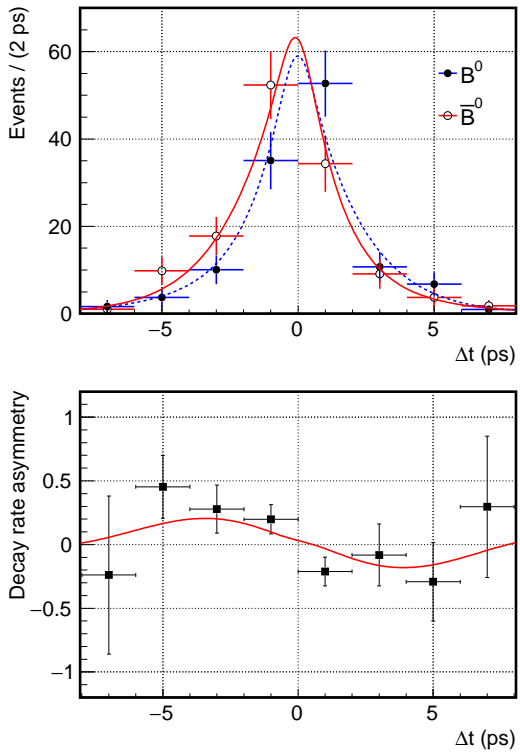}
\caption{
Background-subtracted $\Delta t$ distribution (top) and asymmetry distribution (bottom) obtained from data.
In the $\Delta t$ distribution graph, the red solid line and open circles represent the fitted curve and data for $\overbar{B}^0$, 
while the blue dashed line and filled circles represent the fitted curve and data for $B^{0}$, respectively.
In the asymmetry distribution graph, the points represent data and the solid line represents the fitted curve.
}
\label{fig_cpfit}
\end{figure}

We determine the $C\!P$-violating parameters \cpa~ and \cps~ by performing a second unbinned maximum-likelihood fit.
The contribution to the likelihood function from the $j$th event is 
\begin{equation} \label{eqn_dt4}
\begin{split}
\mathcal{P}_j & = (1-f_{\rm ol}) \times \left [ f^{\rm sig}_j \left \{ \int d(\Delta t') R(\Delta t_j - \Delta t') \times  \right. \right. \\
& \left. \left. \mathcal{P}_{\rm sig} (\Delta t',q_j)  \vphantom{\int} \right \}  + (1-f^{\rm sig}_{j}) \mathcal{P}_{\rm bkg} (\Delta t_j)\right ]  \\
&   +  f_{\rm ol} \mathcal{P}_{\rm ol} (\Delta t_j), 
\end{split}
\end{equation}
where $R(\Delta t)$ is a resolution function.
The resolution function consists of three components: 
detector resolutions for $B_{C\!P}$ and $B_{\rm tag}$,
nonprimary track effects for $B_{\rm tag}$,
and a kinematical approximation due to the difference in the lab momentum of $B_{C\!P}$ and $B_{\rm tag}$ owing to non-zero CM momentum~\cite{vertexres}. 
In Eq.~(\ref{eqn_dt4}), $f_{\rm ol}$ and $\mathcal{P}_{\rm ol}$ are the fraction and
 PDF, respectively, of outlier events for a very long tail shape in the $\Delta t$ distribution, 
 and $f_{j}^{\rm sig}$ is the signal fraction obtained from the signal extraction.
$\mathcal{P}_{\rm bkg}$ is the $\Delta t$ distribution of background events. 
This PDF is the sum of a $\delta$-function and an exponential function, both convolved with a background resolution function.
The parameters of these functions are determined by fitting events in a data sideband region, 
defined as 0.15 GeV $ < |\Delta E| <$ 0.20 GeV or $M_{\rm bc} < 5.26$ GeV$/c^2$.
The unbinned maximum-likelihood fit in the signal region is used to determine the $C\!P$ violation parameters, 
where the world average values are used for $\tau_{B^0}$ and $\Delta m_d$~\cite{PDG}.
The measured \cps ~and \cpa ~are $-0.71 \pm 0.23$ and $0.12 \pm 0.16$,
 respectively, where the uncertainties are statistical.
The background-subtracted $\Delta t$ and asymmetry distributions are shown in Fig.~\ref{fig_cpfit}~\cite{ccbar}.

The estimated systematic uncertainties are summarized in Table~\ref{tab_sys}.
The systematic uncertainty is calculated by varying the fixed parameters in the fit for \cps ~and \cpa, and all uncertainties are summed in quadrature.
For inputs obtained from data and MC, we use $1 \sigma$ and $2 \sigma$ variations, respectively.
\begin{table}[htb]
\caption{Systematic uncertainties}
\label{tab_sys}
\begin{tabular}
 {@{\hspace{0.5cm}}l@{\hspace{0.5cm}}|@{\hspace{0.5cm}}c@{\hspace{0.5cm}}|@{\hspace{0.5cm}}c@{\hspace{0.5cm}}}
\hline \hline
Source & $\delta \mathcal{S}$ & $\delta \mathcal{A}$ \\
\hline
Vertex reconstruction          & 0.031  &  0.038    \\         
Flavor tagging                 & 0.002  &  0.004    \\         
Resolution function            & 0.016  &  0.014    \\         
Physics parameters             & 0.004  &  0.001    \\         
Fit bias                       & 0.012  &  0.009    \\         
Signal fraction                & 0.024  &  0.021    \\         
Background $\Delta t$ shape    & 0.016  &  0.001    \\         
SVD misalignment               &  0.004 &  0.005    \\      
$\Delta z$ bias                &  0.002 &  0.004    \\      
Tag-side interference          & 0.001  &  0.008    \\         
\hline                           
Total                          & 0.047  &  0.047    \\        
\hline \hline
\end{tabular}
\end{table}

The systematic uncertainty on the vertex reconstruction is determined by varying 
 the $x$-$y$ plane smearing parameter for the IP profile,
 charged track requirements for the $B_{\rm tag}$ vertex reconstruction,
 criteria to discard poorly reconstructed vertices for measurement of $C\!P$ violation, and
 correction of helix parameter errors for vertexing.
The systematic uncertainties due to the parameters $w$ and $\Delta w$ are estimated by varying these parameters by their uncertainties.
We vary each resolution function parameter by its uncertainty.
For physics parameters, we calculate differences in \cps ~and \cpa ~
 by varying world average values of $\tau_{B^0}$ and $\Delta m_{B^0}$.
For the systematic uncertainty on the fit bias, we measure $C\!P$ violation parameters using
 the signal MC events, mixed signal MC with continuum toy MC events,
 and mixed signal toy MC with continuum toy MC events at the ratio expected from data.
The larger of the difference between the input \cps ~(\cpa) and the output \cps ~(\cpa)
 and the statistical error on the output \cps ~(\cpa) is considered
 as the uncertainty due to the fit bias.
The systematic uncertainties due to the signal fraction and background $\Delta t$ shape are obtained by varying the parameter $f^{\rm sig}_{j}$ and $\mathcal{P}_{\rm bkg}$ in Eq.~(\ref{eqn_dt4}), respectively.
For possible SVD misalignment, $\Delta z$ bias,
 and tag-side interference, we quote the systematic uncertainties obtained
 from a study of a large control sample of $B \to (c\overbar{c})K_S^0$ decays~\cite{ccbar}. 

\begin{figure}[hbt!]
  \centering
  \includegraphics[width=0.44\textwidth]{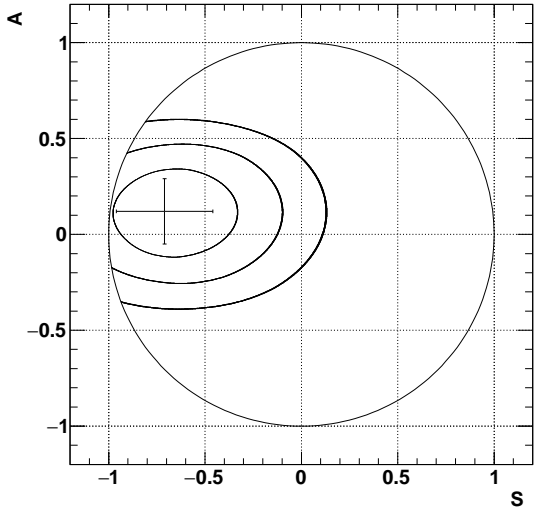}
  \caption{
  Confidence contour for \cps~and \cpa.
  The contours represent 1$\sigma$, 2$\sigma$, and 3$\sigma$ from inner to outer. 
  The point with error bars represents the measured values of \cps~and \cpa.
  The circle is the physical region defined by $\mathcal{S}^2 + \mathcal{A}^2 \leq 1$.
  }
\label{fig_FD}
\end{figure}

The significance, taking both statistical and systematic uncertainties into account, is calculated using a two-dimensional Feldman-Cousins approach~\cite{FD}.
The significance of $C\!P$ violation is determined to be 2.5$\sigma$ away from $(0,0)$ as depicted in Fig.~\ref{fig_FD}, which shows the two-dimensional confidence contour in the \cps~and \cpa~ plane.

In summary, we have studied time-dependent $C\!P$ violation in \bks ~decays using 
 the final data set containing $772 \times 10^6$ $B\overbar{B}$ collected at the Belle experiment.
The NN methods for $K^{0}_{S}$ selection and background suppression and improved vertex reconstruction, along with increased statistics, result in a more precise measurement than the previous Belle one. 
The measured values of \cps~and \cpa~are
\begin{eqnarray}
\mathcal{S} &=& -0.71 \pm 0.23 \rm{(stat)} \pm 0.05 \rm{(syst)}, \nonumber \\
\mathcal{A} &=& 0.12 \pm 0.16 \rm{(stat)} \pm 0.05 \rm{(syst)}. \nonumber
\end{eqnarray}
The results are consistent with the world average value of $-\sin 2 \phi_1$ ($-0.70$)
~\cite{HFLV} as well as with the SM prediction.
These result supersede our previous measurements in Ref.~\cite{b03ksbelle}.
\vskip 1mm
We thank the KEKB group for the excellent operation of the
  accelerator; the KEK cryogenics group for the efficient
  operation of the solenoid; and the KEK computer group, and the Pacific Northwest National
  Laboratory (PNNL) Environmental Molecular Sciences Laboratory (EMSL)
  computing group for strong computing support; and the National
  Institute of Informatics, and Science Information NETwork 5 (SINET5) for
  valuable network support.  We acknowledge support from
  the Ministry of Education, Culture, Sports, Science, and
  Technology (MEXT) of Japan, the Japan Society for the 
  Promotion of Science (JSPS), and the Tau-Lepton Physics 
  Research Center of Nagoya University; 
  the Australian Research Council including grants
  DP180102629, 
  DP170102389, 
  DP170102204, 
  DP150103061, 
  FT130100303; 
  Austrian Science Fund (FWF);
  the National Natural Science Foundation of China under Contracts
  No.~11435013,  
  No.~11475187,  
  No.~11521505,  
  No.~11575017,  
  No.~11675166,  
  No.~11705209;  
  Key Research Program of Frontier Sciences, Chinese Academy of Sciences (CAS), Grant No.~QYZDJ-SSW-SLH011; 
  the  CAS Center for Excellence in Particle Physics (CCEPP); 
  the Shanghai Pujiang Program under Grant No.~18PJ1401000;  
  the Ministry of Education, Youth and Sports of the Czech
  Republic under Contract No.~LTT17020;
  the Carl Zeiss Foundation, the Deutsche Forschungsgemeinschaft, the
  Excellence Cluster Universe, and the VolkswagenStiftung;
  the Department of Science and Technology of India; 
  the Istituto Nazionale di Fisica Nucleare of Italy; 
  National Research Foundation (NRF) of Korea Grant
  Nos.~2016R1\-D1A1B\-01010135, 2016R1\-D1A1B\-02012900, 2018R1\-A2B\-3003643,
  2018R1\-A6A1A\-06024970, 2018R1\-D1A1B\-07047294, 2019K1\-A3A7A\-09033840,
  2019R1\-I1A3A\-01058933, 2019R1\-A6A3A\-01096585; 
Radiation Science Research Institute, Foreign Large-size Research Facility Application Supporting project, the Global Science Experimental Data Hub Center of the Korea Institute of Science and Technology Information and KREONET/GLORIAD;
the Polish Ministry of Science and Higher Education and 
the National Science Center;
the Ministry of Science and Higher Education of the Russian Federation, Agreement 14.W03.31.0026; 
University of Tabuk research grants
S-1440-0321, S-0256-1438, and S-0280-1439 (Saudi Arabia);
the Slovenian Research Agency;
Ikerbasque, Basque Foundation for Science, Spain;
the Swiss National Science Foundation; 
the Ministry of Education and the Ministry of Science and Technology of Taiwan;
and the United States Department of Energy and the National Science Foundation.


\begin{thebibliography}{99}
\bibitem{phi1notation}
Another naming convention, $\beta(=\phi_1)$, is also used in the literature.
\end{thebibliography}
\end{document}